\documentclass[]{interact}

\usepackage{epstopdf}
\usepackage[caption=false]{subfig}

\usepackage[numbers,sort&compress]{natbib}
\bibpunct[, ]{[}{]}{,}{n}{,}{,}
\renewcommand\bibfont{\fontsize{10}{12}\selectfont}

\theoremstyle{plain}

\theoremstyle{definition}

\theoremstyle{remark}

\begin{document}

\articletype{ARTICLE - arXiv.org}

\title{Discrete Bilal distribution with right-censored data}


\author{
	\name{Bruno Caparroz Lopes de Freitas\textsuperscript{a}, Jorge Alberto Achcar\textsuperscript{b}, Marcos Vinicius de Oliveira Peres\textsuperscript{b}, Edson Zangiacomi Martinez\textsuperscript{b}\footnote{Corresponding author: Ribeir\~{a}o Preto Medical School, Av. Bandeirantes 3900, University of S\~{a}o Paulo (USP), Ribeir\~{a}o Preto, 14049-900, Brazil. E-mail: edson@fmrp.usp.br}}
	\affil{\textsuperscript{a}State University of Maring\'{a}, Master Program in Biostatistics, Maring\'{a}, Brazil; \textsuperscript{b}Ribeir\~{a}o Preto Medical School, University of S\~{a}o Paulo (USP), Ribeir\~{a}o Preto, Brazil}
}

\maketitle

\begin{abstract}
This paper presents inferences for the discrete Bilal (DB) distribution introduced by Altun et al. (2020). We consider parameter estimation for DB distribution in the presence of randomly right-censored data. We use maximum likelihood and Bayesian methods for the estimation of the model parameters. We also consider the inclusion of a cure fraction in the model. The usefulness of the proposed model was illustrated with three examples considering real datasets. These applications suggested that the model based on DB distribution performs at least as good as some other traditional discrete models as the DsFx-I, discrete Lindley, discrete Rayleigh, and discrete Burr-Hatke distributions. R codes are provided in an appendix at the end of the paper so that reader can carry out their own analysis. 
\end{abstract}


\section{Introduction}

Survival analysis is one of the statistical techniques most commonly encountered in the medical literature (\citeauthor{Flynn2012}, \citeyear{Flynn2012}). These methods are applied when the time until the occurrence of an event is the object of interest. Examples in medical research include the time to respond to treatment, relapse-free survival time, time to death, time to device failure, and time to regain mobility (\citeauthor{Myers2007}, \citeyear{Myers2007}). The Kaplan-Meier plots, log-rank tests, and Cox (proportional hazards) regression model are the most widely used survival analysis techniques in medical studies (\citeauthor{LeRademacher2021}, \citeyear{LeRademacher2021}). As an alternative to the traditional proportional hazard model, parametric models have become popular in the last decades. The parametric models assume that the time-to-event variable follows a known probability distribution, such as Weibull, gamma, or the log-normal distributions. Among the discrete distributions proposed in the statistical literature to model time-to-event data, we have the discrete Weibull distribution (\citeauthor{Nakagawa1975}, \citeyear{Nakagawa1975}), the discrete Lindley distribution (\citeauthor{Gomez2011}, \citeyear{Gomez2011}), the exponentiated discrete Weibull distribution (\citeauthor{Nekoukhou2015}, \citeyear{Nekoukhou2015}; \citeauthor{Cardial2020}, \citeyear{Cardial2020}; \citeauthor{Freitas2021}, \citeyear{Freitas2021}), the discrete generalized Rayleigh distribution (\citeauthor{Alamatsaz2016}, \citeyear{Alamatsaz2016}), and the discrete Sushila distribution (\citeauthor{Oliveira2019}, \citeyear{Oliveira2019}).

Let $X$ be a random variable denoting a survival time, and let $x$ be an observation of $X$. The continuous Bilal distribution introduced by \citeauthor{AbdElrahman2013} (\citeyear{AbdElrahman2013}) has a probability density function (pdf) given by%
\[
f_{X}(x)=\frac{6}{\theta }e^{-\frac{2x}{\theta }}\left( 1-e^{-\frac{x}{\theta }%
}\right) ,\text{ \ }x\geq 0,\text{ }\theta >0
\]%
and probability accumulated distribution function given by%
\[
F_{X}(x)=1-e^{-\frac{2x}{\theta }}\left( 3-2e^{-\frac{x}{\theta }}\right) .
\]
The survival function, that is, the probability that an individual survives at least until time $x$, is given by $S_{X}(x)=1-F_{X}(x)$. The author named this distribution as Bilal since this is his youngest son's name (\citeauthor{AbdElrahman2013}, \citeyear{AbdElrahman2013}). Classical and Bayesian approaches to find an estimated value for the parameter $\theta $ of a Bilal distribution based on a given type-2 right censoring sample are given by \citeauthor{AbdElrahman2017} (\citeyear{AbdElrahman2017}). Generalizations of the Bilal distribution are found in the works of \citeauthor{AbdElrahman2019} (\citeyear{AbdElrahman2019}) and \citeauthor{Shi2021} (\citeyear{Shi2021}).

\begin{figure}[h]
	\centering
	\includegraphics[width=1\linewidth]{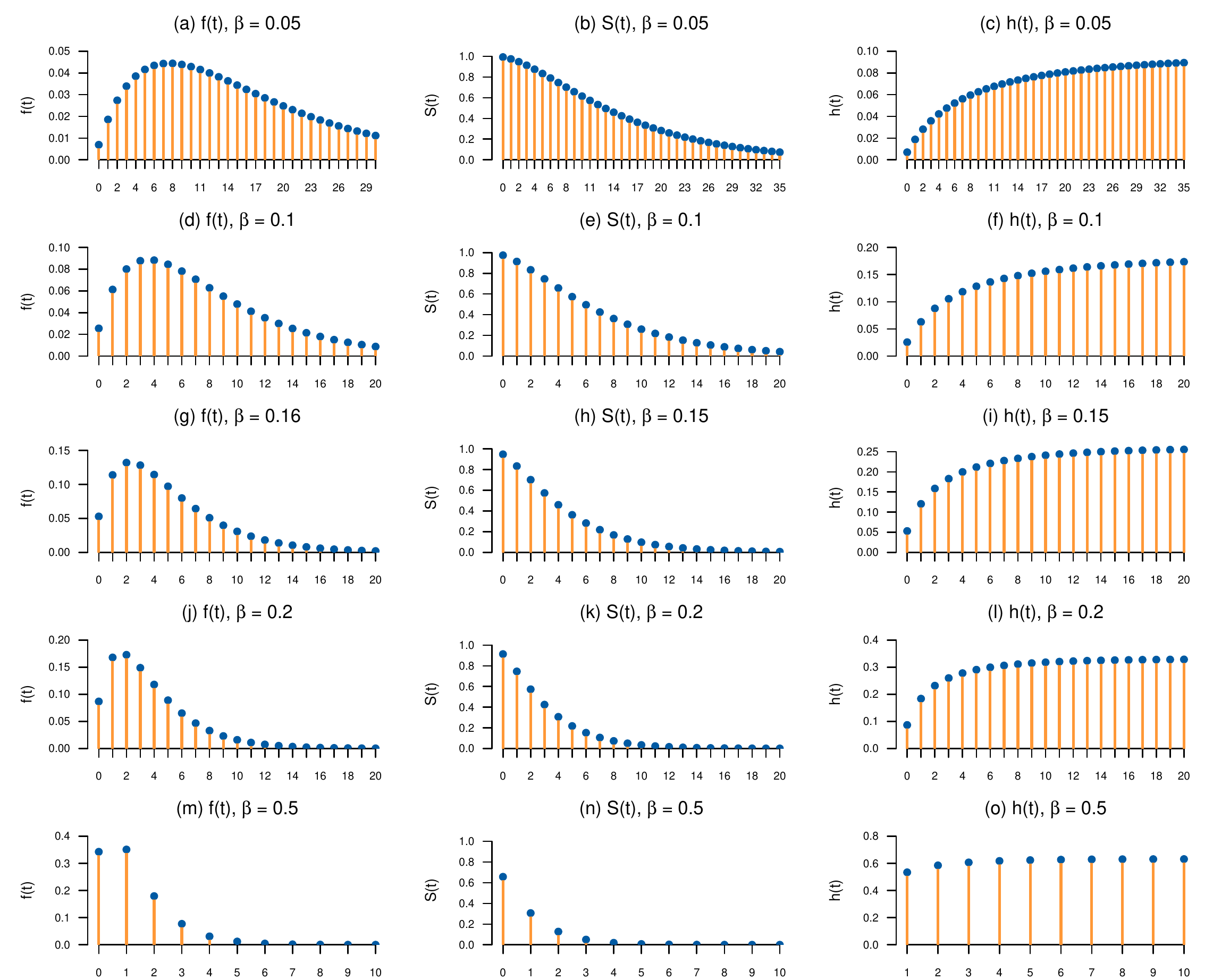}
	\caption[figDES]{The pmf, survival function and hazard function of DB($\beta$) for different values of $\beta$.\bigskip}
	\label{fig:fig01}
\end{figure}

To obtain a discrete version of the Bilal distribution, \citeauthor{Altun2020} (\citeyear{Altun2020}) considered that the random variable $T$ has probability mass function (pmf) given by%
\begin{equation}
	P(T=t)=P(t \leq X \leq t+1)=S_{X}(t)-S_{X}(t+1),\text{ \ }t\in 
	\mathbb{N}
	^{0},  \label{PX}
\end{equation}%
where $X$ is the underlying continuous random variable, $T=[X]$ (the largest integer less than or equal to $X$), and $S_{X}(x)=P(X>x)$ (see Methodology-IV in the article by \citeauthor{Chakraborty2015} (\citeyear{Chakraborty2015})). Thus, replacing $S_{X}(t)$ by $e^{-\frac{2t}{\theta }}\left( 3-2e^{-\frac{t}{\theta }}\right) $ and $S_{X}(t+1)$ by $e^{-\frac{2(t+1)}{\theta }}\left( 3-2e^{-\frac{%
		t+1}{\theta }}\right) $ in the expression (\ref{PX}), we have
\begin{equation}
	P(T=t)=e^{-\frac{2t}{\theta }}\left( 3-2e^{-\frac{t}{\theta }}\right) -e^{-%
		\frac{2(t+1)}{\theta }}\left( 3-2e^{-\frac{t+1}{\theta }}\right) .
	\label{PX2}
\end{equation}
Let us assume the parameter transformation $p=e^{-\frac{1}{\theta }}$, $0<p<1$. From (\ref{PX2}) and, following the notation of \citeauthor{Altun2020} (\citeyear{Altun2020}), the pmf of the discrete Bilal distribution is given by%
\begin{equation}
	f(t)=P(T=t)=2(p^{3}-1)p^{3t}-3(p^{2}-1)p^{2t},\text{ }t\in 
	\mathbb{N}
	^{0},  \label{f(t)}
\end{equation}%
where $0<p<1$. The corresponding probability accumulated distribution function is given by%
\[
F(t)=P(T\leq t)=1-(3-2p^{t+1})p^{2(t+1)},\text{ }t\in 
\mathbb{N}
^{0},
\]%
and the survival function is thus given by%
\[
S(t)=1-F(t)=P(T>t)=(3-2p^{t+1})p^{2(t+1)},\text{ }t\in 
\mathbb{N}
^{0}.
\]
To simplify the obtaining of estimators for the parameters of the discrete Bilal distribution, we consider the reparameterization $%
p=e^{-\beta }$, where $\beta >0$. Thus, the pmf is given by
\[
f(t)=P(T=t)=2(e^{-3\beta }-1)e^{-3\beta t}-3(e^{-2\beta }-1)e^{-2\beta t},%
\text{ }t\in 
\mathbb{N}
^{0},
\]
where the corresponding probability accumulated distribution function is given by
\[
F(t)=P(T\leq t)=1-\left[ 3-2e^{-\beta \left( t+1\right) }\right] e^{-2\beta
	(t+1)},\text{ }t\in 
\mathbb{N}
^{0},
\]
and the survival function is
\[
S(t)=\left[ 3-2e^{-\beta \left( t+1\right) }\right] e^{-2\beta (t+1)},\text{ 
}t\in 
\mathbb{N}
^{0}.
\]
The corresponding hazard function is given by%
\[
h(t)=P(\left. T=t\right\vert T\geq t)=\frac{P(T=t)}{P(T\geq t)}=\frac{f(t)}{%
	S(t-1)}=\frac{2(e^{-3\beta }-1)e^{-\beta t}-3(e^{-2\beta }-1)}{3-2e^{-\beta
		t}}.
\]
\citeauthor{Altun2020} (\citeyear{Altun2020}) showed that the mean and the variance of a random variable $T$ that follows a
discrete Bilal distribution with parameter $\beta$ are respectively given by%
\[
E(T)=\frac{e^{-2\beta }\left( e^{-2\beta }+e^{-\beta }+3\right) }{\left(
	e^{-2\beta }+e^{-\beta }+1\right) \left( 1-e^{-2\beta }\right) }
\]
and%
\[
Var(T)=\frac{e^{-2\beta }\left( 3e^{-4\beta }+4e^{-3\beta
	}-e^{-2\beta }+4e^{-\beta }+3\right) }{\left( e^{-2\beta }+e^{-\beta
	}+1\right) ^{2}\left( e^{-2\beta }-1\right) ^{2}}\text{.}
\]
Let us denote a discrete Bilal distribution with parameter $\beta$ as $DB(\beta)$. Figure \ref{fig:fig01} presents graphs of the pmf, survival function, and hazard function of the discrete Bilal (DB) distribution considering different values for $\beta $. We can note that the DB distribution has an increasing hazard function.

The novelty of the present article consists in introducing the DB distribution to model lifetime data in the presence of right-censored time-to-event data. We also
consider the inclusion of a cure fraction in the model. This paper is organized as follows. Section \ref{MET} presents the maximum-likelihood (ML)
estimation for the parameter of the DB distribution based on complete and censored data. ML estimation in the presence of censored data and a cure fraction is also discussed in this section. In addition, Section \ref{MET} presents a Bayesian framework for the model. Three examples considering real data from the medical literature are used in Section \ref{EX} to illustrate the usefulness of this model to a broad range of problems. Finally, in Section \ref{CR}, some concluding remarks are presented. The computational codes used in this article are provided in the Appendix.
\section{Methods} \label{MET}
\subsection{Maximum likelihood estimation for complete data}
After some algebra we can see that the equation (\ref{f(t)}) is equivalent to%
\begin{equation}
	f(t)=P(T=t)=p^{2t}(p-1)\left[ 2p^{t}(p^{2}+p+1)-3p-3\right] .  \label{ft2}
\end{equation}%
Let $T_1$,...,$T_n$ be a random sample of failure times from a DB survival distribution. Considering $p=e^{-\beta }$ and the expression (\ref{ft2}), the likelihood function for the parameter $\beta $ is given by%
\[
L(\left. \beta \right\vert \mathbf{t})=\prod\limits_{i=1}^{n}e^{-2\beta
	t_{i}}(e^{-\beta }-1)\left[ 2e^{-\beta t_{i}}(e^{-2\beta }+e^{-\beta
}+1)-3e^{-\beta }-3\right] ,
\]%
and the corresponding log-likelihood function is given by%
\[
\ell (\left. \beta \right\vert \mathbf{t})=-2\beta
\sum\limits_{i=1}^{n}t_{i}+n\log (e^{-\beta }-1)+\sum\limits_{i=1}^{n}\log %
\left[ 2e^{-\beta t_{i}}(e^{-2\beta }+e^{-\beta }+1)-3e^{-\beta }-3\right] .
\]
By deriving the log-likelihood function with respect to $\beta $, we have
the following equation:%
\begin{equation}
	\frac{d\ell }{d\beta }=-2\sum\limits_{i=1}^{n}t_{i}-\frac{ne^{-\beta }}{%
		e^{-\beta }-1}-\sum\limits_{i=1}^{n}\frac{2e^{-\beta (t_{i}-1)}\left(
		1+2e^{-\beta }\right) +2t_{i}e^{-\beta t_{i}}\left( e^{-2\beta }+e^{-\beta
		}+1\right) -3e^{-\beta }}{2e^{-\beta t_{i}}\left( e^{-2\beta }+e^{-\beta
		}+1\right) -3e^{-\beta }-3}. \label{der1}
\end{equation}%
The maximum-likelihood (ML) estimator $\widehat{\beta }_{ML}$ for $\beta$ is obtained by equating the right-hand side of (\ref{der1}) to zero and solving for $\beta$. Nevertheless, the resulting expression has not a closed-form solution and so numerical methods are needed to find the ML estimate for $\beta$. In this article, we use the  maxLik package in R program to obtain the ML estimate of the parameter $\beta$ (\citeauthor{Henningsen2011}, \citeyear{Henningsen2011}). A confidence interval for $\beta$ can be constructed from the asymptotic normality of the ML estimate considering large sample sizes, given by%
\[
\widehat{\beta }_{ML}\sim N\left( \beta ,\widehat{Var}(\widehat{\beta }%
_{ML})\right) ,
\]%
where, in the single-parameter case, $\widehat{Var}(\widehat{\beta }_{ML})$ is the estimated variance for $\widehat{\beta }_{ML}$. Therefore, an
approximate $100(1-\upsilon )\%$ Wald-type confidence interval (CI) for $\beta $ is given by%
\[
\widehat{\beta }_{ML}\mp z_{\upsilon /2}\sqrt{\widehat{Var}(\widehat{\beta }%
	_{ML})},
\]%
where $z_{\upsilon }$ denotes the upper $\upsilon $-th percentile of the
standard normal distribution. The asymptotic variance of a ML estimator can be estimated
by the negative of the inverse of the second derivative of the log-likelihood function evaluated at $\widehat{\beta }_{ML}$. Thus, we have%
\begin{equation}
	\frac{d^{2}\ell }{d\beta ^{2}}=-\frac{ne^{-\beta }}{\left( e^{-\beta
		}-1\right) ^{2}}+\sum_{i=1}^{n}\frac{A_{i}}{\left[ 2e^{-\beta
			t_{i}}(e^{-2\beta }+e^{-\beta }+1)-3e^{-\beta }-3\right] ^{2}},  \label{d2}
\end{equation}%
where%
\begin{eqnarray*}
	A_{i} &=&-2e^{-\beta \left( t_{i}-1\right) }(3-2e^{-\beta t_{i}}+6e^{-2\beta}+12e^{-\beta }) +2t_{i}e^{-\beta \left( t_{i}-1\right) }(3+9e^{-\beta }-6e^{-2\beta}+4e^{-\beta t_{i}}) \\
	&&+ 6t_{i}e^{-\beta t_{i}}\left( 2e^{-2\beta }-e^{-2\beta }-e^{-\beta}+t_{i}\right) +3e^{-\beta }\left( 2e^{-\beta t_{i}}-2e^{-2\beta }e^{-\beta	t_{i}}-3\right)  \\
	&&+ 4e^{-2\beta }(6t_{i}-e^{-\beta }+6t_{i}e^{-\beta }+4t_{i}e^{-2\beta }-4)	+6t_{i}^{2}e^{-\beta (1+t_{i})}\left( 2e^{-\beta }+e^{-2\beta }+2\right) .
\end{eqnarray*}
The negative second derivative of the log-likelihood function is the observed information denoted by $I_{n}(\beta)$, that is,%
\[
I_{n}(\beta) = -\frac{d^{2}\ell }{d\beta ^{2}}.
\]%
The expected value of $I_{n}(\beta)$, say $i_{n}(\beta)$, is called the expected Fisher information. The asymptotic variance of $\widehat{\beta }_{ML}$ is given by the inverse of the expected information evaluated at the ML estimate of $\beta$, that is,%
\[
\widehat{Var}(\widehat{\beta }_{ML}) = \left[i_{n}(\widehat{\beta }_{ML}) \right]^{-1}.
\]%
\subsection{Maximum-likelihood estimation in presence of censored data}
Considering a random sample $(t_{i},d_{i})$ of size $n$, $i=1,\cdots,n$, the contribution of the $i$th individual to the likelihood function is given by%
\[
L_{i}=\left[ f(t_{i})\right] ^{d_{i}}\left[ S(t_{i})\right] ^{1-d_{i}},
\]
where $d_{i}$ is a censoring indicator variable, that is, $d_{i} = 1$ for an observed survival time and $d_{i} = 0$ for a right-censored survival time. Assuming the data with a DB distribution, the likelihood function for the parameter $\beta $ is given by%
\begin{eqnarray*}
	L(\left. \beta \right\vert \mathbf{t},\mathbf{d})
	&=&\prod\limits_{i=1}^{n}e^{-2\beta t_{i}d_{i}}(e^{-\beta }-1)^{d_{i}}\left[
	2e^{-\beta t_{i}}(e^{-2\beta }+e^{-\beta }+1)-3e^{-\beta }-3\right] ^{d_{i}}
	\\
	&&\times \left[ 3-2e^{-\beta \left( t_{i}+1\right) }\right] ^{\left(
		1-d_{i}\right) }e^{-2\beta (t_{i}+1)\left( 1-d_{i}\right) },
\end{eqnarray*}%
and the corresponding log-likelihood function is%
\begin{eqnarray*}
	\ell (\left. \beta \right\vert \mathbf{t},\mathbf{d}) &=&2\beta
	\sum\limits_{i=1}^{n}t_{i}d_{i}+\log (e^{-\beta
	}-1)\sum\limits_{i=1}^{n}d_{i} \\
    &&+\sum\limits_{i=1}^{n}d_{i}\log \left[
	2e^{-\beta t_{i}}(e^{-2\beta }+e^{-\beta }+1)-3e^{-\beta }-3\right]  \\
	&&+\sum\limits_{i=1}^{n}(1-d_{i})\log \left[ 3-2e^{-\beta \left(
		t_{i}+1\right) }\right] -2\beta \sum\limits_{i=1}^{n}(t_{i}+1)\left(
	1-d_{i}\right) .
\end{eqnarray*}
By deriving the log-likelihood function with respect to $\beta$, we have%
\begin{eqnarray*}
	\frac{d\ell }{d\beta } &=&2\sum\limits_{i=1}^{n}t_{i}d_{i}-\frac{e^{-\beta }%
	}{e^{-\beta }-1}\sum_{i=1}^{n}d_{i} \\
	&&-\sum_{i=1}^{n}d_{i}\frac{2e^{-\beta t_{i}}\left( e^{-\beta }+2e^{-2\beta
		}+t_{i}+t_{i}e^{-2\beta }+t_{i}e^{-\beta }\right) -3e^{-\beta }}{2e^{-\beta
			t_{i}}(e^{-2\beta }+e^{-\beta }+1)-3e^{-\beta }-3} \\
	&&-2\sum\limits_{i=1}^{n}(1-d_{i})\frac{(t_{i}+1)e^{-\beta \left(
			t_{i}+1\right) }}{2e^{-\beta \left( t_{i}+1\right) }-3}-2\sum%
	\limits_{i=1}^{n}(t_{i}+1)\left( 1-d_{i}\right) .
\end{eqnarray*}
Setting this expression equal to zero, we get the corresponding score equation whose numerical solution leads to the ML estimator. The second derivative of the log-likelihood function with respect to $\beta$ is given by%
\begin{eqnarray*}
	\frac{d^{2}\ell }{d\beta ^{2}} &=&-\sum_{i=1}^{n}d_{i}\frac{e^{-\beta }}{%
		\left( e^{-\beta }-1\right) ^{2}}-\sum_{i=1}^{n}d_{i}\frac{B_{i}}{\left[
		2e^{-\beta t_{i}}(e^{-2\beta }+e^{-\beta }+1)-3e^{-\beta }-3\right] ^{2}} \\
	&&-6\sum\limits_{i=1}^{n}(1-d_{i})\frac{e^{-\beta \left( t_{i}+1\right)
		}\left( t_{i}+1\right) ^{2}}{\left[ 2e^{-\beta \left( t_{i}+1\right) }-3%
		\right] ^{2}},
\end{eqnarray*}%
where%
\begin{eqnarray*}
	B_{i} &=&6t_{i}^{2}e^{-\beta t_{i}}\left( e^{-3\beta }+2e^{-2\beta
	}+2e^{-\beta }e^{-\beta t_{i}}+1\right)  \\
	&&+12t_{i}e^{-\beta t_{i}}e^{-2\beta }\left( e^{-\beta }+2\right)
	+6e^{-\beta t_{i}}e^{-\beta }\left( e^{-2\beta }+4e^{-\beta }+2\right)  \\
	&&-4e^{-2\beta t_{i}}e^{-\beta }\left( e^{-2\beta }+4e^{-\beta }+1\right)
	-9e^{-\beta }.
\end{eqnarray*}
Approximated CI for $\beta$ also could be obtained based on the asymptotical normality of the ML estimate for $\beta$ in similar way as described in the previous subsection.

Given that the time-to-event variable is discrete, randomized quantile
residuals can be used to test model adequacy (\citeauthor{Dunn1996}, \citeyear{Dunn1996}). These
residuals are given by $r_{i}=\Phi ^{-1}(u_{i})$, $i=1,...,n$, where $\Phi
(\cdot )$ is the standard normal distribution function and $u_{i}$ is an
uniform random variable on the interval $\left[ F(t_{i}-1,\widehat{\beta }%
_{ML}),F(t_{i},\widehat{\beta }_{ML})\right] $ if $d_{i}=1$ and $\left[
F(t_{i},\widehat{\beta }_{ML}),1\right] $ if $d_{i}=0$. Let us consider that 
$F(t_{i},\widehat{\beta }_{ML})$ is the cumulative distribution function of
the DB distribution based on the ML estimate for $\beta $. The randomized quantile
residuals are expected to follow the standard normal distribution if the
model is correct. Hence, a normal Q-Q plot can be used to visually check the
normality of these residuals.

\subsection{Maximum-likelihood estimation including censored data and a cure fraction} \label{cure}
A fundamental characteristic of the traditional survival analysis methods is that the survival function $S(t)$ converges to zero when the
time variable tends to infinity. In the applications of survival methods to medical data, this implies assuming that all individuals under study are susceptible to the event of interest. However, there are situations where this assumption is not satisfied (\citeauthor{Othus2012}, \citeyear{Othus2012}). For example, in randomized trials evaluating the efficacy of a treatment for a disease of interest, it is possible that some patients may be cured of the disease due to the treatment under study. If the event of interest is death due to this disease, these patients are no longer subject to this event. The presence of cured individuals in a data set is usually suggested by a of stable plateau at the right tail of the Kaplan–Meier non-parametric estimator of the survival function, with heavy censoring in this portion of the plot (\citeauthor{Corbiere2009}, \citeyear{Corbiere2009}). Different parametric and non-parametric approaches that consider the presence of immune individuals have been proposed in the literature (\citeauthor{Maller1996}, \citeyear{Maller1996}; \citeauthor{Amico2018}, \citeyear{Amico2018}; \citeauthor{Peng2021}, \citeyear{Peng2021}). These approaches include the mixture model, which explicitly includes a parameter accounting for a fraction of immune individuals (\citeauthor{Lambert2007}, \citeyear{Lambert2007}; \citeauthor{Martinez2013}, \citeyear{Martinez2013}). This model assumes that the probability of observing a survival time greater than or equal to some fixed value $t$ is given by the survival function
\[
S(t)=\eta +(1-\eta )S_{0}(t),
\]
where $\eta$ is the proportion of immune, cured or not susceptible individuals, and $S_{0}(t)$ is the baseline survival function for the susceptible individuals (\citeauthor{Farewell1982}, \citeyear{Farewell1982}). Considering a random sample $(t_{i},d_{i})$ of size $n$, $i=1,\cdots,n$, the contribution of the $i$th individual to the likelihood function is given by
\[
L_{i}=\left[ f\left( t_{i}\right) \right] ^{d_{i}}\left[ S\left(
t_{i}\right) \right] ^{1-d_{i}}=\left[ (1-\eta )f_{0}\left( t_{i}\right) %
\right] ^{d_{i}}\left[ \eta +(1-\eta )S_{0}(t)\right] ^{1-d_{i}},
\]
where $d_{i}$ is a binary censoring indicator variable and $f_{0}(t)$ is the corresponding baseline probability function. Assuming the mixture model based on the DB distribution, the likelihood function for $\beta $ and $\eta $ is given by%
\begin{eqnarray*}
	L(\left. \beta ,\eta \right\vert \mathbf{t},\mathbf{d})
	&=&\prod\limits_{i=1}^{n}(1-\eta )^{d_{i}}e^{-2\beta d_{i}t_{i}}(e^{-\beta
	}-1)^{d_{i}}\left[ 2e^{-\beta t_{i}}(e^{-2\beta }+e^{-\beta }+1)-3e^{-\beta
	}-3\right] ^{d_{i}} \\
	&&\times \left\{ \eta +(1-\eta )\left[ 3-2e^{-\beta \left( t_{i}+1\right) }%
	\right] e^{-2\beta (t_{i}+1)}\right\} ^{1-d_{i}} .
\end{eqnarray*}
The log-likelihood function in this case is 
\begin{eqnarray*}
	\ell (\left. \beta ,\eta \right\vert \mathbf{t},\mathbf{d})
	&=&\sum\limits_{i=1}^{n}d_{i}\log (1-\eta )-2\beta
	\sum\limits_{i=1}^{n}d_{i}t_{i}+\sum\limits_{i=1}^{n}d_{i}\log (e^{-\beta
	}-1) \\
	&&+\sum\limits_{i=1}^{n}d_{i}\log \left[ 2e^{-\beta t_{i}}(e^{-2\beta
	}+e^{-\beta }+1)-3e^{-\beta }-3\right]  \\
	&&+\sum\limits_{i=1}^{n}(1-d_{i})\log \left\{ \eta +(1-\eta )\left[
	3-2e^{-\beta \left( t_{i}+1\right) }\right] e^{-2\beta (t_{i}+1)}\right\} .
\end{eqnarray*}
The first derivative of the log-likelihood function with respect to $\beta $ is given by%
\begin{eqnarray*}
	\frac{\partial \ell }{\partial \beta } &=&-2\sum%
	\limits_{i=1}^{n}d_{i}t_{i}-\sum_{i=1}^{n}d_{i}\frac{e^{-\beta }}{e^{-\beta
		}-1} \\
	&&-\sum_{i=1}^{n}d_{i}\frac{2e^{-\beta t_{i}}\left[ t_{i}+e^{-\beta }\left(
		2e^{-\beta }+t_{i}e^{-\beta }+t_{i}+1\right) \right] -3e^{-\beta }}{%
		2e^{-\beta t_{i}}(e^{-2\beta }+e^{-\beta }+1)-3e^{-\beta }-3} \\
	&&+6\left( \eta -1\right) \sum\limits_{i=1}^{n}(1-d_{i})\frac{\left(
		t_{i}+1\right) e^{-2\beta \left( t_{i}+1\right) }\left[ e^{-\beta \left(
			t_{i}+1\right) }-1\right] }{\eta +\left( 1-\eta \right) e^{-2\beta \left(
			t_{i}+1\right) }\left\{ 3+2\left[ e^{-2\beta \left( t_{i}+1\right) }\right]
		^{\frac{1}{2}}\right\} },
\end{eqnarray*}%
and the first derivative of the log-likelihood function with respect to $%
\eta $ is given by%
\[
\frac{\partial \ell }{\partial \eta }=-\frac{1}{1-\eta }\sum_{i=1}^{n}d_{i}+%
\sum\limits_{i=1}^{n}(1-d_{i})\frac{\left[ 2e^{-\beta \left( t_{i}+1\right)
	}+1\right] \left[ e^{-\beta \left( t_{i}+1\right) }-1\right] ^{2}}{\eta
	+\left( 1-\eta \right) e^{-2\beta \left( t_{i}+1\right) }\left\{ 3+2\left[
	e^{-2\beta \left( t_{i}+1\right) }\right] ^{\frac{1}{2}}\right\} }.
\]
Setting these expressions equal to zero and solving them simultaneously we get the ML estimators of the parameters $\beta$ and $\eta$. Although we
cannot obtain explicit expressions for the ML estimators for these parameters, they can be estimated numerically using iterative algorithms such as the Newton-Raphson method and its variants.

The second partial derivatives of the ML function are given as follows:%
\[
\frac{\partial ^{2}\ell }{\partial \beta ^{2}}=-\sum_{i=1}^{n}d_{i}\frac{%
	e^{-\beta }}{\left( e^{-\beta }-1\right) ^{2}}+6\left( \eta -1\right)
\sum\limits_{i=1}^{n}(1-d_{i})\left( t_{i}+1\right) ^{2}e^{-2\beta \left(
	t_{i}+1\right) }\frac{C_{i}}{D_{i}},
\]%
where%
\[
C_{i}=\eta \left( 2-3e^{^{-\beta \left( t_{i}+1\right) }}\right) +5\left(\eta -1\right) e^{-3\beta \left( t_{i}+1\right) }
\]%
and%
\begin{eqnarray*}
	D_{i} &=&\eta ^{2}+6\eta \left( 1-\eta \right) e^{-2\beta \left(t_{i}+1\right) }+4\eta e^{-3\beta \left( t_{i}+1\right) }+9\left( 1-2\eta
	\right) e^{-4\beta \left( t_{i}+1\right) }+4\left( 1-\eta \right)^{2}e^{-6\beta \left( t_{i}+1\right) } \\
	&&+12\left( 1-2\eta \right) e^{-4\beta \left( t_{i}+1\right) }e^{^{-\beta\left( t_{i}+1\right) }}+9\eta ^{2}e^{-4\beta \left( t_{i}+1\right) }-4\eta
	^{2}e^{-3\beta \left( t_{i}+1\right) }+12\eta ^{2}e^{-5\beta \left(t_{i}+1\right) },
\end{eqnarray*}
\[
\frac{\partial ^{2}\ell }{\partial \eta ^{2}}=\frac{1}{\left( \eta
	-1\right) ^{2}}\sum_{i=1}^{n}d_{i}+\sum\limits_{i=1}^{n}(1-d_{i})\frac{%
	\left[ 2e^{-\beta \left( t_{i}+1\right) }+1\right] \left[ e^{-\beta \left(
		t_{i}+1\right) }-1\right] ^{2}\left[ 3e^{-2\beta \left( t_{i}+1\right)
	}+2e^{-3\beta \left( t_{i}+1\right) }-1\right] }{E_{i}},
\]%
where%
\begin{eqnarray*}
	E_{i} &=&\eta ^{2}+6\eta \left( 1-\eta \right) e^{-2\beta \left(
		t_{i}+1\right) }+4\eta \left( 1-\eta ^{2}\right) e^{-3\beta \left(t_{i}+1\right) } \\
	&&+\left( 1-\eta \right) ^{2}\left[ 9e^{-4\beta \left( t_{i}+1\right)}+12e^{-5\beta (t_{i}+1)}+4e^{-6\beta \left( t_{i}+1\right) }\right],
\end{eqnarray*}
and%
\[
\frac{\partial ^{2}\ell }{\partial \beta \partial \eta }=6\sum%
\limits_{i=1}^{n}(1-d_{i})\frac{\left( t_{i}+1\right) e^{-2\beta \left(
		t_{i}+1\right) }\left[ e^{-\beta \left( t_{i}+1\right) }-1\right] }{F_{i}},
\]%
where%
\[
F_{i}=\eta ^{2}+2\eta \left( 1-\eta \right) \left[ 3e^{-2\beta \left(t_{i}+1\right) }+2e^{-3\beta \left( t_{i}+1\right) }\right] +\left( 1-\eta
\right) ^{2}\left[ 9e^{-4\beta \left( t_{i}+1\right) }+12e^{-5\beta \left(t_{i}+1\right) }+4e^{-6\beta \left( t_{i}+1\right) }\right] .
\]

The asymptotic multivariate normal distribution of the ML estimators $%
\widehat{\beta }_{ML}$ and $\widehat{\eta }_{ML}$ is denoted by%
\[
\left[ 
\begin{array}{c}
	\widehat{\beta }_{ML} \\ 
	\widehat{\eta }_{ML}%
\end{array}%
\right] \sim N\left( \left[ 
\begin{array}{c}
	\beta  \\ 
	\eta 
\end{array}%
\right] ,\left[ 
\begin{array}{cc}
	V_{11} & V_{12} \\ 
	V_{21} & V_{22}%
\end{array}%
\right] \right) ,
\]%
where $V_{11}$ is the variance of $\widehat{\beta }_{ML}$, $V_{22}$ is the
variance of $\widehat{\eta }_{ML}$, and $V_{12}=V_{12}$ is the covariance
between $\widehat{\beta }_{ML}$ and $\widehat{\eta }_{ML}$. Approximate $100(1-\upsilon )\%$ Wald-type CIs for $\beta$ and $\eta$ are, respectively, given by%
\[
\widehat{\beta }_{ML}\mp z_{\upsilon /2}\sqrt{\widehat{Var}(\widehat{\beta }%
	_{ML})}\text{ \ \ and \ \ }\widehat{\eta }_{ML}\mp z_{\upsilon /2}\sqrt{%
	\widehat{Var}(\widehat{\eta }_{ML})},
\]%
where $z_{\upsilon }$ denotes the upper $\upsilon $-th percentile of the
standard normal distribution. The asymptotic variances of the ML estimators
are given by the elements of the inverse of the Fisher's information matrix.
The expected information matrix is given by%
\[
i_{n}(\beta ,\eta )=\left[ 
\begin{array}{cc}
	-E\left( \frac{\partial ^{2}\ell }{\partial \beta ^{2}}\right)  & -E\left( 
	\frac{\partial ^{2}\ell }{\partial \beta \partial \eta }\right)  \\ 
	-E\left( \frac{\partial ^{2}\ell }{\partial \beta \partial \eta }\right)  & 
	-E\left( \frac{\partial ^{2}\ell }{\partial \eta ^{2}}\right) 
\end{array}%
\right] ,
\]
where the derivatives are provided above. R code for implementing this procedure is presented in the Appendix.

\subsection{Bayesian analysis} \label{Bayes}
The Bayesian approach is an alternative to the ML estimation of the parameters. In the Bayesian inference it is necessary to specify a prior distribution for each
unknown parameter (\citeauthor{Gelman2013}, \citeyear{Gelman2013}). From the Bayes' theorem, the posterior distribution of a particular parameter under the model
specification is proportional to its prior distribution multiplied by the likelihood of the data. Considering the discrete Bilal distribution, we can assume a gamma prior distribution to the $\beta$ parameter. That is, $\beta \sim Gamma(a_{\beta},b_{\beta})$, where $a_{\beta}$ and $b_{\beta}$ are known hyperparameters and $Gamma(a,b)$ denotes a gamma distribution with mean $a/b$ and variance $a/b^2$. In the case of the model in the presence of a cure fraction, we can consider a prior  $\eta \sim Beta(a_{\eta},b_{\eta})$ to obtain a Bayesian estimate of $\eta$, where $a_{\eta}$ and $b_{\eta}$ are known hyperparameters and $Beta(a,b)$ denotes a beta distribution with mean $a/(a+b)$ and variance $ab/[(a+b)^{2}(a+b+1)]$. Further, we assumed prior independence between the parameters $\beta$ and $\eta$. 

In this article, posterior summaries of interest were obtained using standard Markov-chain Monte Carlo (MCMC) procedures as the Gibbs sampling. The simulation algorithm generated 1,005,000 samples of the joint posterior distribution of interest with a burn-in phase of 5,000 simulated samples to eliminate the effect of the initial values in the iterative procedure and considered a thinning interval of size 200 to have approximately independent samples. The Bayes estimates of the
parameters were obtained as the mean of samples drawn from the joint posterior distribution. The convergence of the simulated sequences was monitored by using traceplots and the Geweke diagnostic (\citeauthor{Geweke1992}, \citeyear{Geweke1992}). The Geweke convergence diagnostic is based on a z score that compares the difference in the two means of non-overlapping sections of a simulated Markov chain, divided by the asymptotic standard error of the difference. This z score asymptotically follows a standardized normal distribution, so we obtain convergence for a chain if its correspondent absolute z score is less than 1.96. Posterior summaries of interest were obtained using the MCMCpack package of the R software (\citeauthor{Martin2006}, \citeyear{Martin2006}). See Appendix for details about the R code used in this article.

\section{Examples} \label{EX}
In this section, we illustrate the estimation procedure proposed here with three examples from the literature. We compare the fits of the discrete Bilal distribution with some competitive models such as DsFx-I (\citeauthor{Eliwa2021}, \citeyear{Eliwa2021}), discrete Lindley (\citeauthor{Gomez2011}, \citeyear{Gomez2011}), discrete Rayleigh (\citeauthor{Roy2004}, \citeyear{Roy2004}), and discrete Burr-Hatke (\citeauthor{ElMorshedy2020}, \citeyear{ElMorshedy2020}) distributions. All these distributions have only one parameter to estimate. The fitted models are compared using Akaike and Bayesian information criteria (AIC and BIC). When the sample size is small, many authors have suggested using the corrected AIC (AICC) as an alternative to AIC (\citeauthor{Hurvich1989}, \citeyear{Hurvich1989}).

\begin{figure}[h!]
	\centering
	\includegraphics[width=1\linewidth]{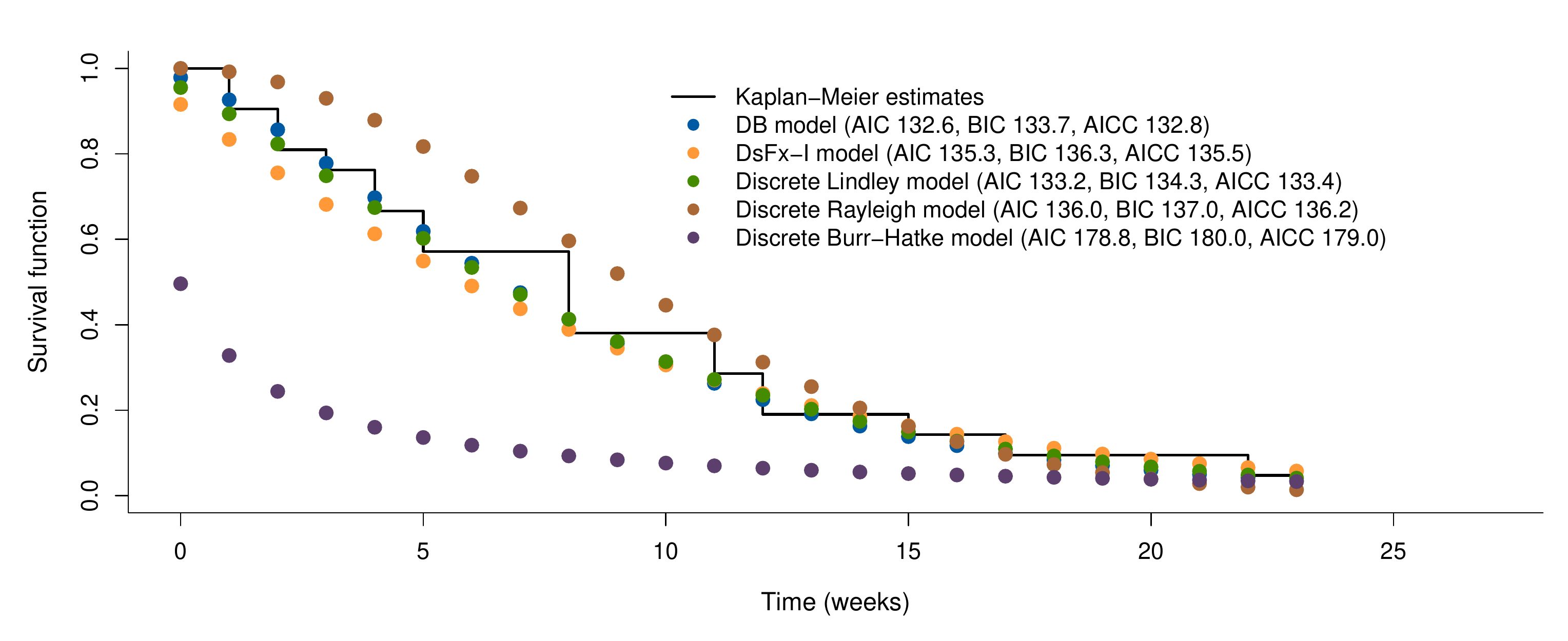}
	\caption[figDES]{Survival function for the acute leukemia patients' data estimated by the Kaplan-Meier method and by using
		the models based on the DB, DsFx-I, discrete Lindley, discrete Rayleigh, and discrete Burr-Hatke distributions.\bigskip}
	\label{fig:FigKMEx01}
\end{figure}
\begin{figure}[h!]
	\centering
	\includegraphics[width=1\linewidth]{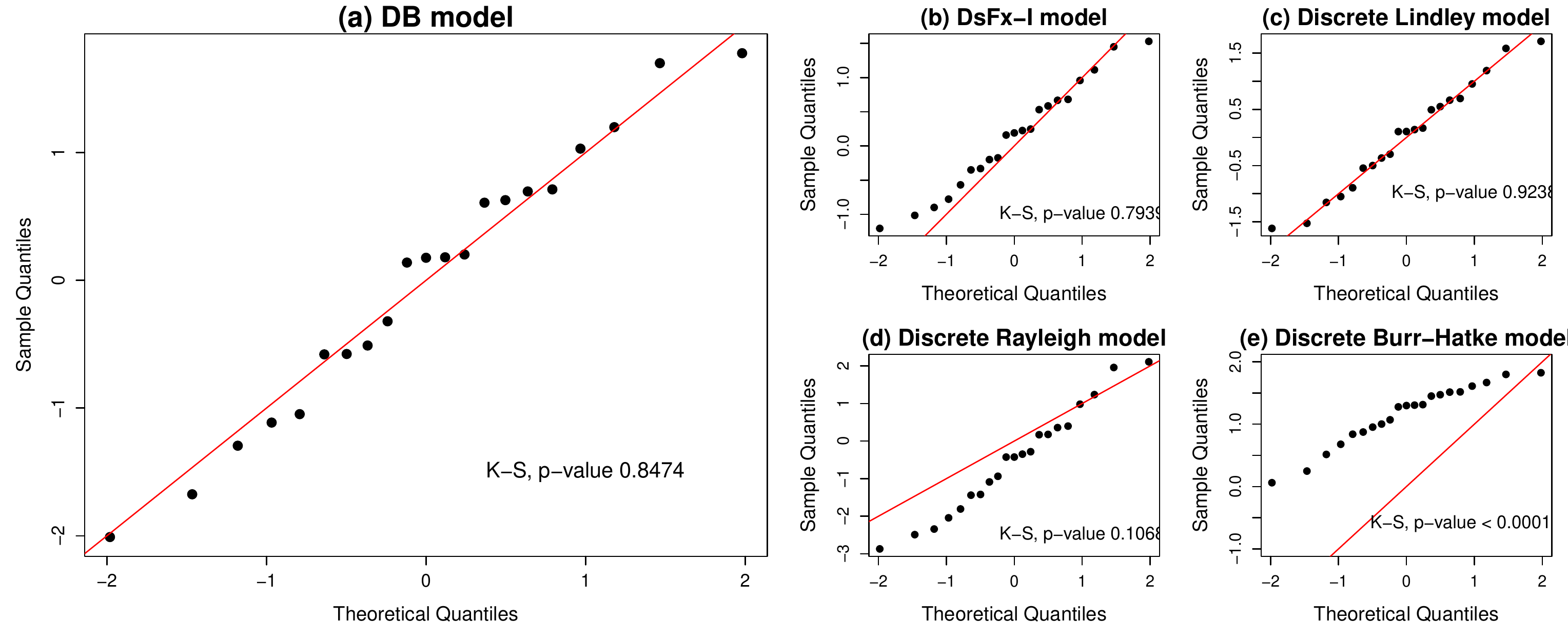}
	\caption[figDES]{Randomized quantile residuals for the models based on the (a) DB, (b) DsFx-I, (c) discrete Lindley, (d) discrete Rayleigh, and (e) discrete Burr-Hatke distributions, fitted to data from patients with acute leukemia.\bigskip}
	\label{fig:FigResidualsEx01}
\end{figure}
\subsection{Patients with acute leukemia}
In this subsection, a numerical example with complete data is presented to illustrate the applicability of the discrete Bilal distribution. A total of $97$ patients 
with acute leukemia participated in a clinical trial investigating the effect of 6-mercaptopurine 
on the duration of steroid-induced remissions (\citeauthor{Freireich1963}, \citeyear{Freireich1963}). The remission times for the $n=21$ patients treated with placebo were 
1, 1, 2, 2, 3, 4, 4, 5, 5, 8, 8, 8, 8, 11, 11, 12, 12, 15, 17, 22, and 23 weeks. Using the maxLik package of the R software, we obtained an ML estimate of $\widehat{\beta }_{ML} = 0.09085$ for the parameter $\beta$ of the discrete Bilal distribution, with a standard error of $0.01431$. An approximate $95\%$ Wald-type CI for $\beta $ is $(0.0628, 0.1189)$. Figure \ref{fig:FigKMEx01} compares the survival function estimated by the Kaplan-Meier method and fitted by parametric models based on the discrete Bilal, DsFx-I, discrete Lindley, discrete Rayleigh, and discrete Burr-Hatke distributions. Figure \ref{fig:FigKMEx01} also shows the corresponding AIC, BIC, and AICC values. Figure \ref{fig:FigResidualsEx01} shows the resulting residual analysis and the corresponding p-values for the Kolmogorov-Smirnov (K-S) test for normality. The model based on the DB distribution appears to fit the data reasonably well, as does the model based on the Lindley distribution. Figures \ref{fig:FigKMEx01} and \ref{fig:FigResidualsEx01} suggest that the discrete Rayleigh and discrete Burr-Hatke distributions do not fit the data well.
\begin{figure}[h!]
	\centering
	\includegraphics[width=1\linewidth]{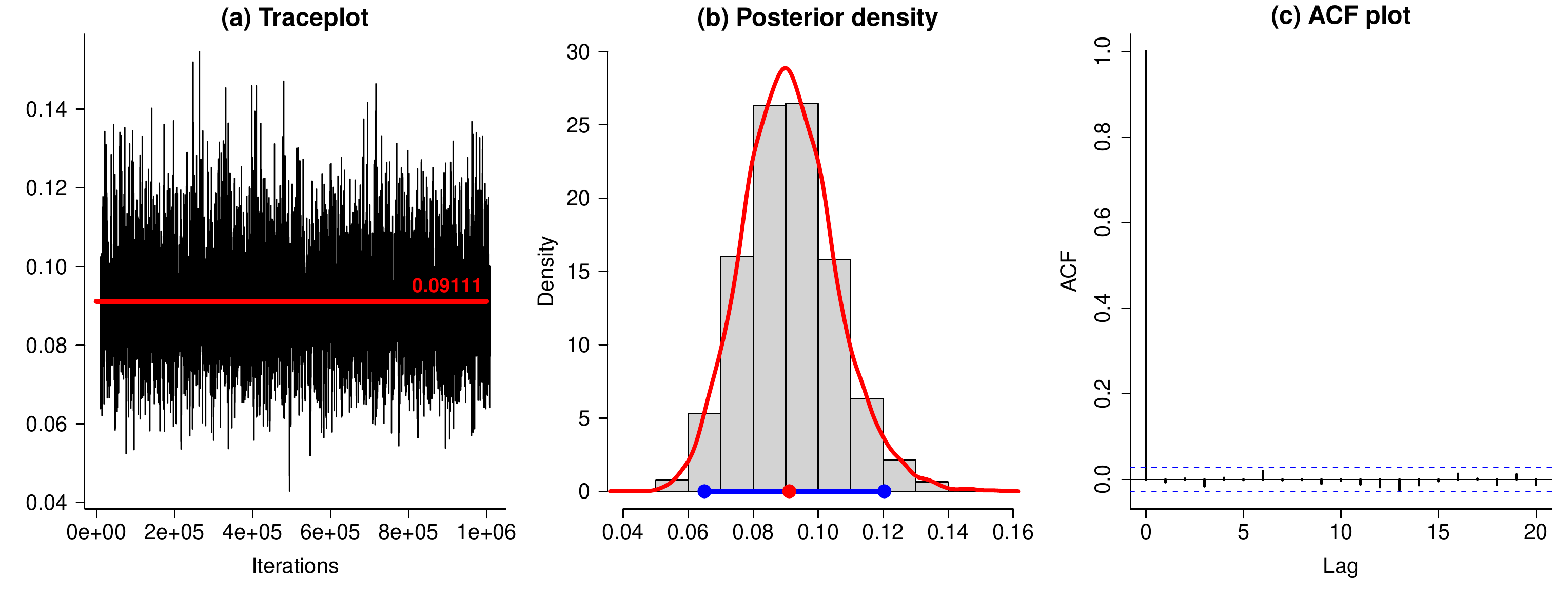}
	\caption[figDES]{Posterior samples for the model parameters of the DB distribution applied to data from patients with acute leukemia. (a) Traceplot of posterior samples, (b) histogram and posterior density with the correspondent $95\%$ HDI (blue line), and (c) auto-correlation function (ACF) plot for the posterior samples of the model parameters.\bigskip}
	\label{fig:FigBayesEx01}
\end{figure}

For a Bayesian data analysis, it was assumed an approximately non-informative gamma prior distribution for the parameter $\beta$ of the DB distribution, that is, $\beta \sim Gamma(0.001,0.001)$. Posterior samples for $\beta$ are described in Figure \ref{fig:FigBayesEx01}. The traceplot shown in panel (a) shows the evolution of the
MCMC draws over the iterations, indicating that the generated samples reached good convergence. The plot of the autocorrelation function (ACF) shows that the posterior samples are uncorrelated (panel (c)). The corresponding Geweke z-score is $0.183$, also suggesting satisfactory convergence of the samples to a stable distribution. The posterior mean for $\beta$ is $\widehat{\beta }_{Bayes} = 0.09111$, and the corresponding $95\%$ HDI (highest density interval) is $(0.0650,  0.1203)$. The  $95\%$ HDI is plotted on the histogram shown in panel (b) of Figure \ref{fig:FigBayesEx01}. We can note that the ML and the Bayesian estimates are fairly close to each other.

\begin{figure}[h]
	\centering
	\includegraphics[width=1\linewidth]{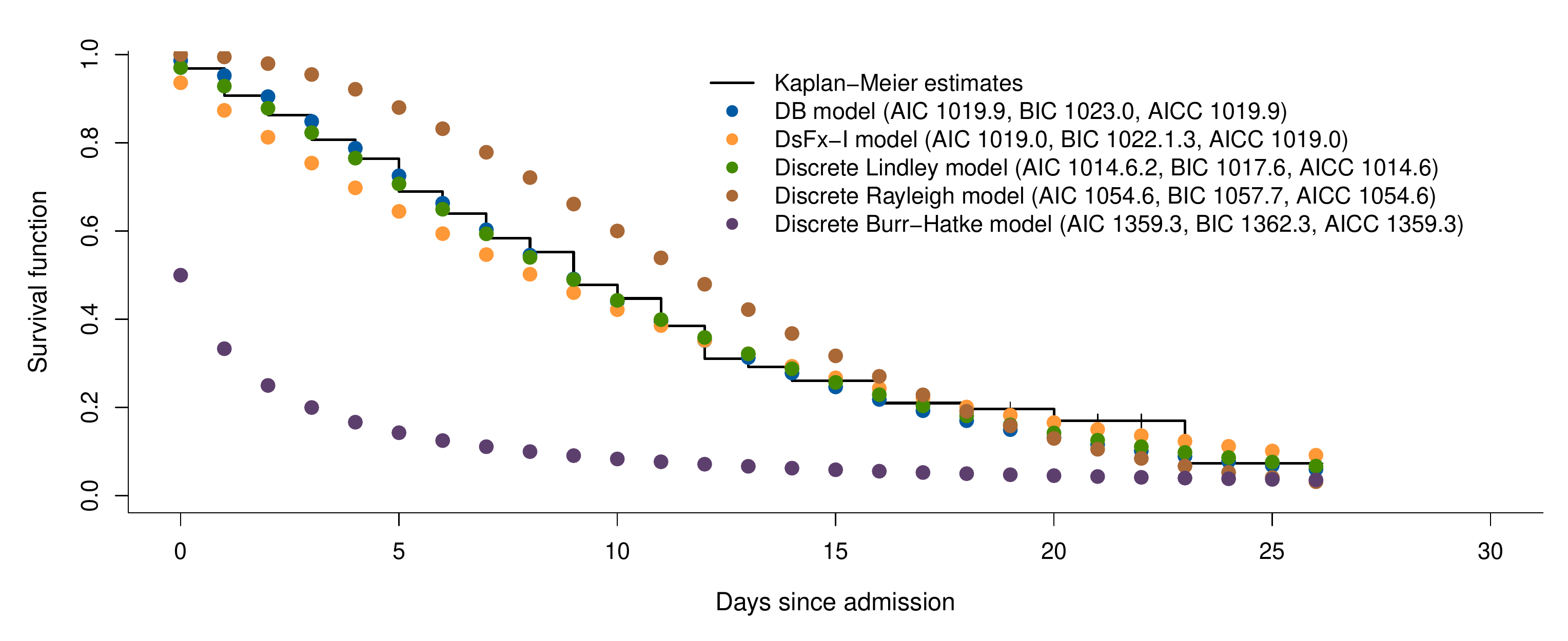}
	\caption[figDES]{Survival function for the COVID-19 patients' data estimated by the Kaplan-Meier method and by using
		the models based on the DB, DsFx-I, discrete Lindley, discrete Rayleigh, and discrete Burr-Hatke distributions.\bigskip}
	\label{fig:FigKMEx02}
\end{figure}

\begin{figure}[h]
	\centering
	\includegraphics[width=1\linewidth]{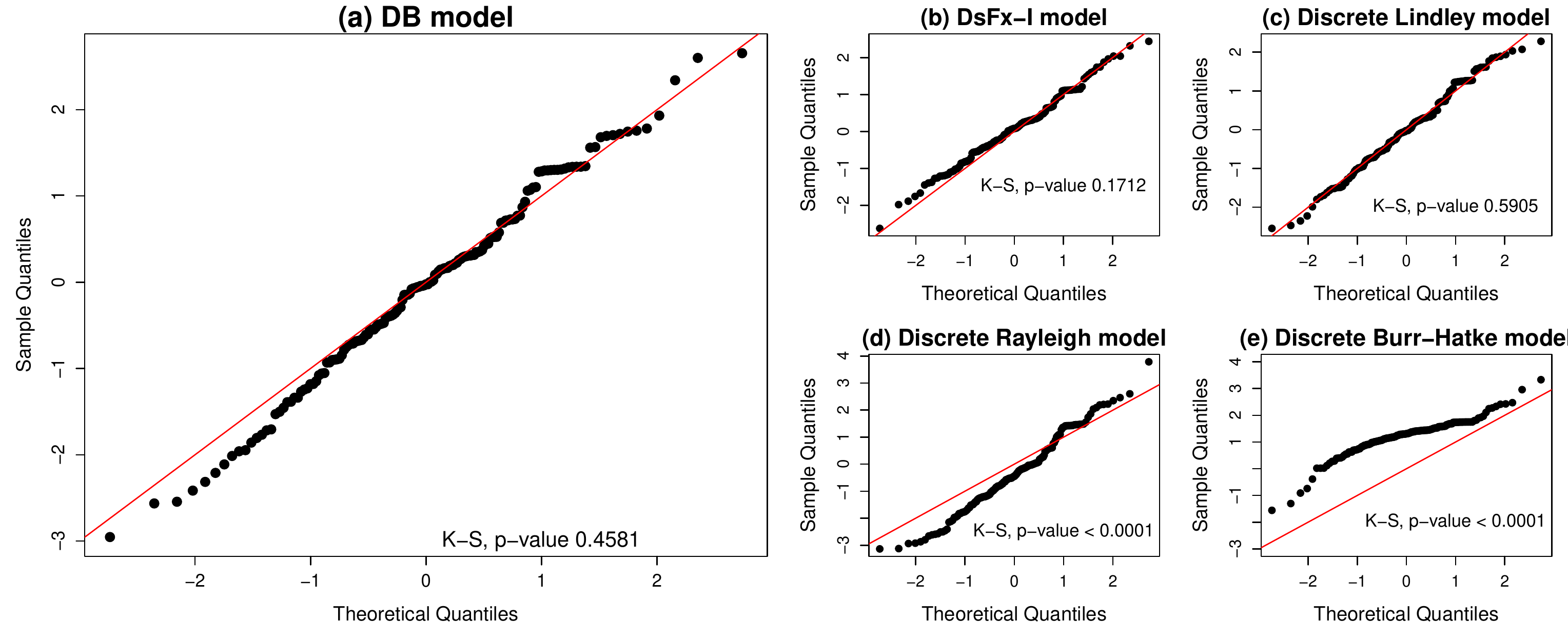}
	\caption[figDES]{Randomized quantile residuals for the models based on the (a) DB, (b) DsFx-I, (c) discrete Lindley, (d) discrete Rayleigh, and (e) discrete Burr-Hatke distributions, fitted to data from hospitalized patients with COVID-19.\bigskip}
	\label{fig:FigResidualsEx02}
\end{figure}

\subsection{Hospitalized patients with COVID-19}
The study by (\citeauthor{Paranjpe2020}, \citeyear{Paranjpe2020}) assessed the association between administration of in-hospital anticoagulation and survival in a large cohort of hospitalized patients with COVID-19 in the Mount Sinai Health System, New York City. In this example, we consider a subsample of $n=161$ patients who required mechanical ventilation and were not treated with in-hospital systemic anticoagulation. The variable of interest is the time from admission to death, in days. We have $15$ $(9.3\%)$ censored observations. As the data were available in figures and not in numerical form, we used the open-source software WebPlotDigitizer, a web based tool to extract numerical data from images (\citeauthor{Drevon2017}, \citeyear{Drevon2017}; \citeauthor{Rohatgi2020}, \citeyear{Rohatgi2020}). In this example, the ML estimate for the parameter $\beta$ of the discrete Bilal distribution is $\widehat{\beta }_{ML} = 0.07047$ (standard error $0.00413$, $95\%$ Wald-type CI 0.0624 to 0.0786). Figure \ref{fig:FigKMEx02} compares the survival function estimated by the Kaplan-Meier method and fitted by parametric models based on the DB and other distributions. Figure \ref{fig:FigResidualsEx02} shows the randomized quantile residuals from the fitted models. We note that the DB distribution fitted the data as well as the DsFx-I and the discrete Lindley distributions, but the models based on the discrete Rayleigh and discrete Burr-Hatke distributions did not fit the data well.

\begin{figure}[h!]
	\centering
	\includegraphics[width=1\linewidth]{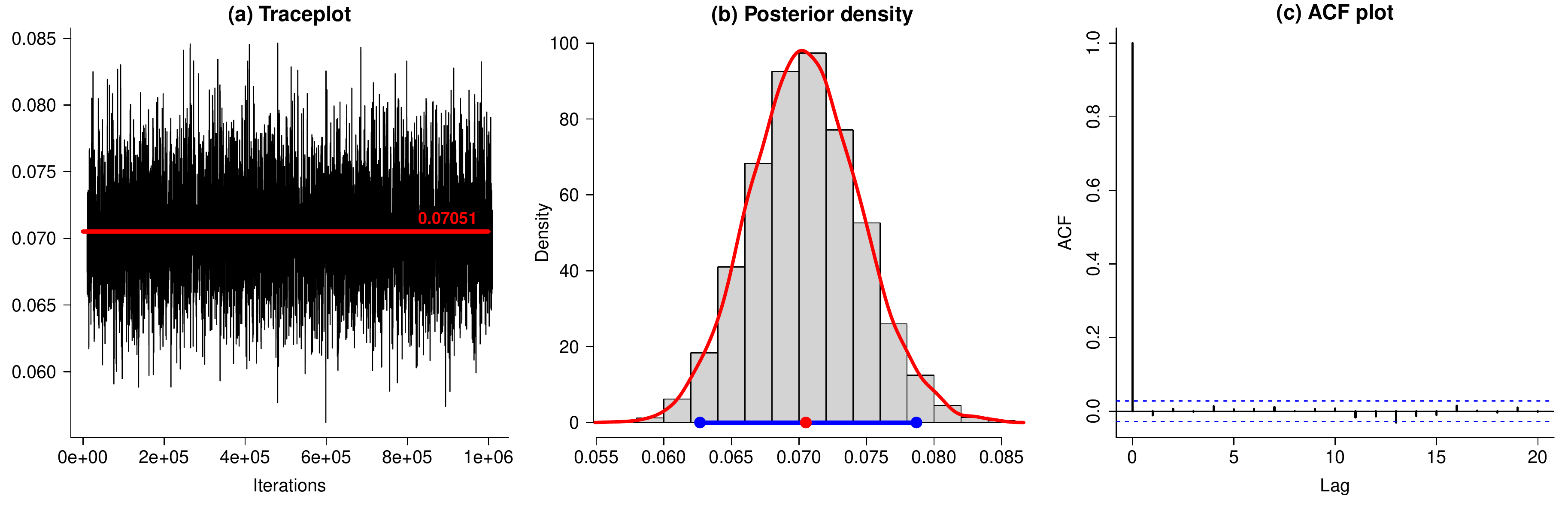}
	\caption[figDES]{Posterior samples for the parameter of the DB distribution applied to data from hospitalized patients with COVID-19. (a) Traceplot of posterior samples, (b) histogram and posterior density with the correspondent $95\%$ HDI (blue line), and (c) auto-correlation function (ACF) plot for the posterior samples of the model parameter.\bigskip}
	\label{fig:FigBayesEx02}
\end{figure}
\begin{figure}[h!]
	\centering
	\includegraphics[width=1\linewidth]{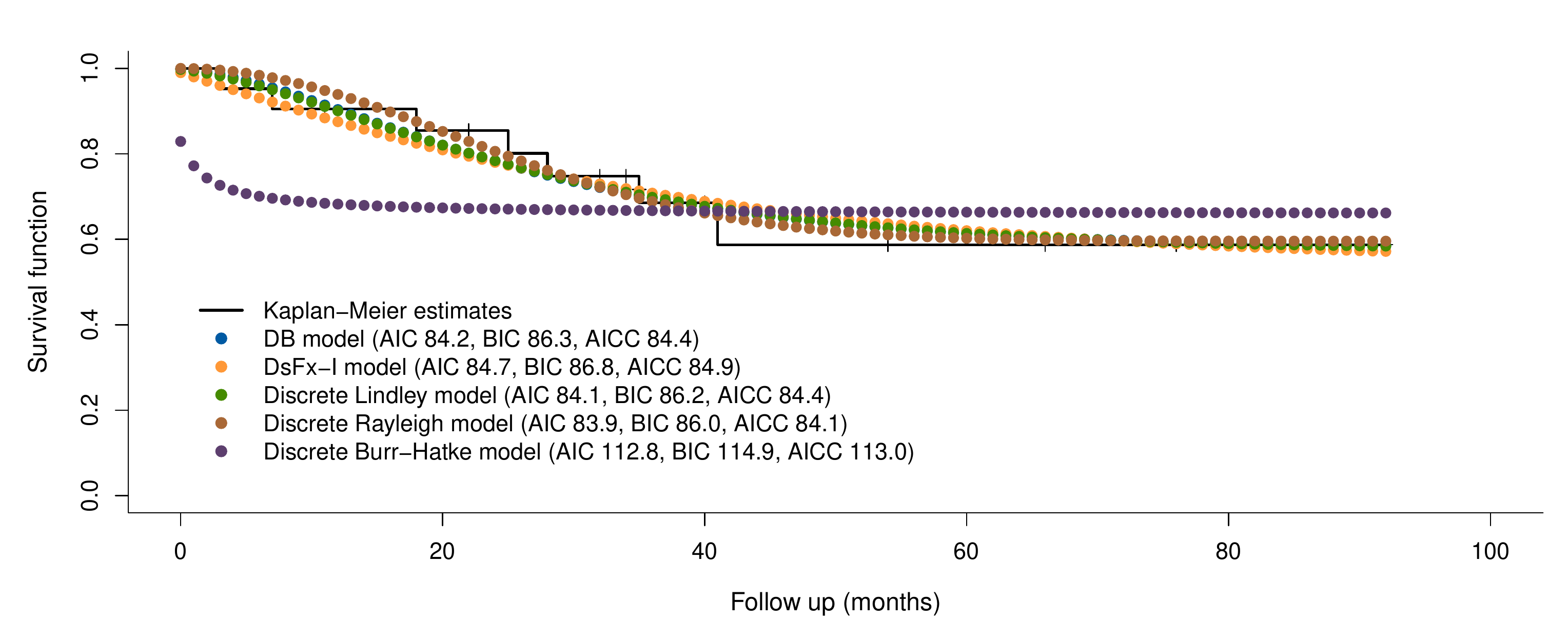}
	\caption[figDES]{Survival function for the pelvic tumors patients' data estimated by the Kaplan-Meier method and by using
		the models based on the DB, DsFx-I, discrete Lindley, discrete Rayleigh, and discrete Burr-Hatke distributions.\bigskip}
	\label{fig:FigKMEx03}
\end{figure}
\begin{figure}[h!]
	\centering
	\includegraphics[width=1\linewidth]{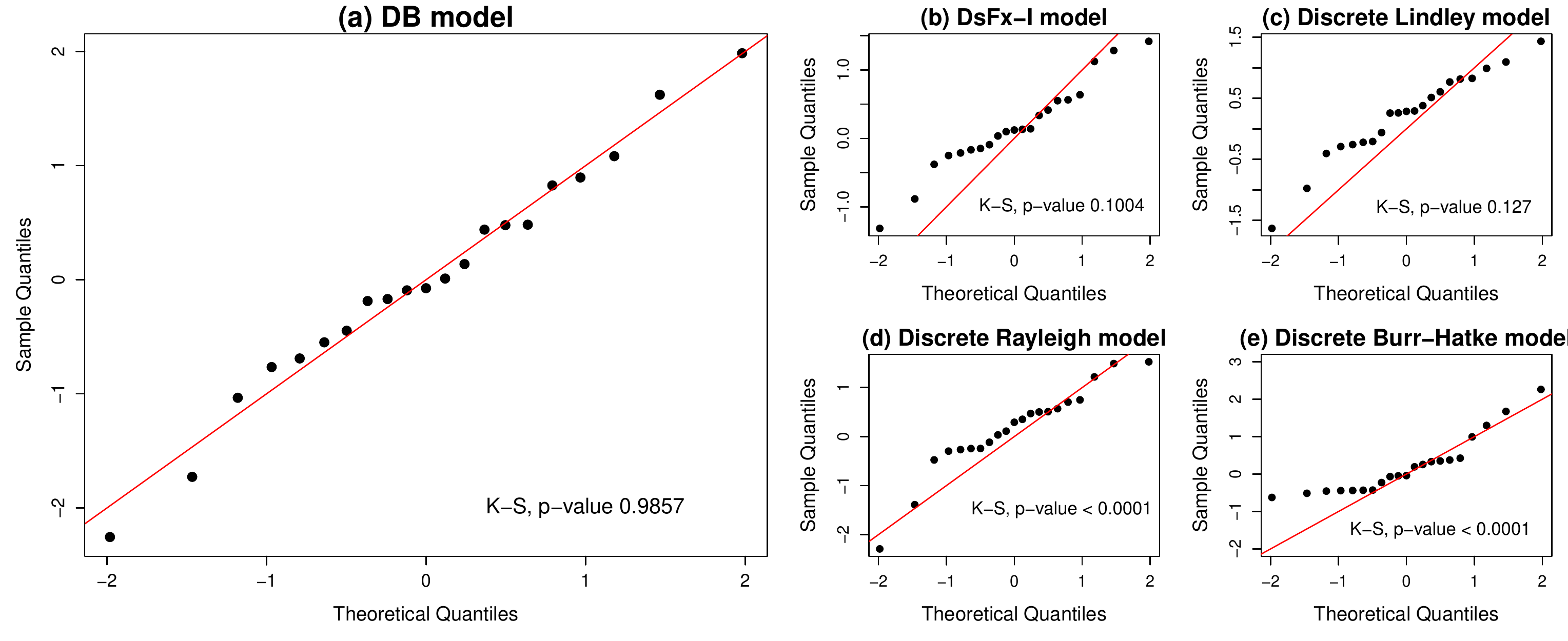}
	\caption[figDES]{Randomized quantile residuals for the models based on the (a) DB, (b) DsFx-I, (c) discrete Lindley, (d) discrete Rayleigh, and (e) discrete Burr-Hatke distributions, fitted to data from patients with pelvic tumors.\bigskip}
	\label{fig:FigResidualsEx03}
\end{figure}
In the Bayesian analysis, as in the previous example, we assumed a gamma prior distribution for the parameter $\beta$ given by $\beta \sim Gamma(0.001,0.001)$. Figure \ref{fig:FigBayesEx02} describes the posterior samples for $\beta$. The traceplot in panel (a) shows that the MCMC algorithm has stabilized, and the corresponding Geweke z-score is $0.033$, also suggesting satisfactory convergence. Panel (b) describes the shape of the posterior distribution and shows the $95\%$ HDI. The posterior mean for $\beta$ is $\widehat{\beta }_{Bayes} = 0.07051$, and the corresponding $95\%$ HDI is $(0.0626, 0.0787)$. This MCMC estimate is closer to the corresponding ML estimate. The ACF plot shows that autocorrelations are not significantly different from zero (panel (c)).

\subsection{Recurrence rates of pelvic tumors with marginal or intracapsular margins}
In this example, we consider a model for survival data with a cure fraction based on the DB distribution. Let us consider the data from a study undertaken at the Musculoskeletal Oncology Center of the First Affiliated Hospital of Sun Yat-Sen University, China, between 2003 and 2013 (\citeauthor{Wang2015}, \citeyear{Wang2015}).
The objective of this study was to evaluate the effectiveness of reconstruction with a modular hemipelvic endoprosthesis after pelvic tumor resection. The recurrence times of pelvic tumors with marginal or intracapsular margins were 3, 7, 11+, 18, 22+, 25, 28, 32+, 34+, 35, 35+, 36+, 40+, 40+, 41, 54+, 66+, 76+, 84+, 88+, and 92+ months, where + denotes a censored observation. By applying the model described in subsection \ref{cure} to these data, we obtained the ML estimates $\widehat{\beta }_{ML} = 0.02859$ (standard error 0.01047, $95\% CI$ 0.0081 to 0.0491) and $\widehat{\eta }_{ML} = 0.57985$ (standard error 0.13965, $95\% CI$ 0.3061 to 0.8536). Figure \ref{fig:FigKMEx03} compares the Kaplan-Meier estimates and the ML estimates of the survival function corresponding to the model based on DB distribution and the concurrent discrete distributions. We can note that the results provided by models based on DB and discrete Lindley distributions are almost identical and are overlapping on the graph.
\begin{figure}[h!]
	\centering
	\includegraphics[width=1\linewidth]{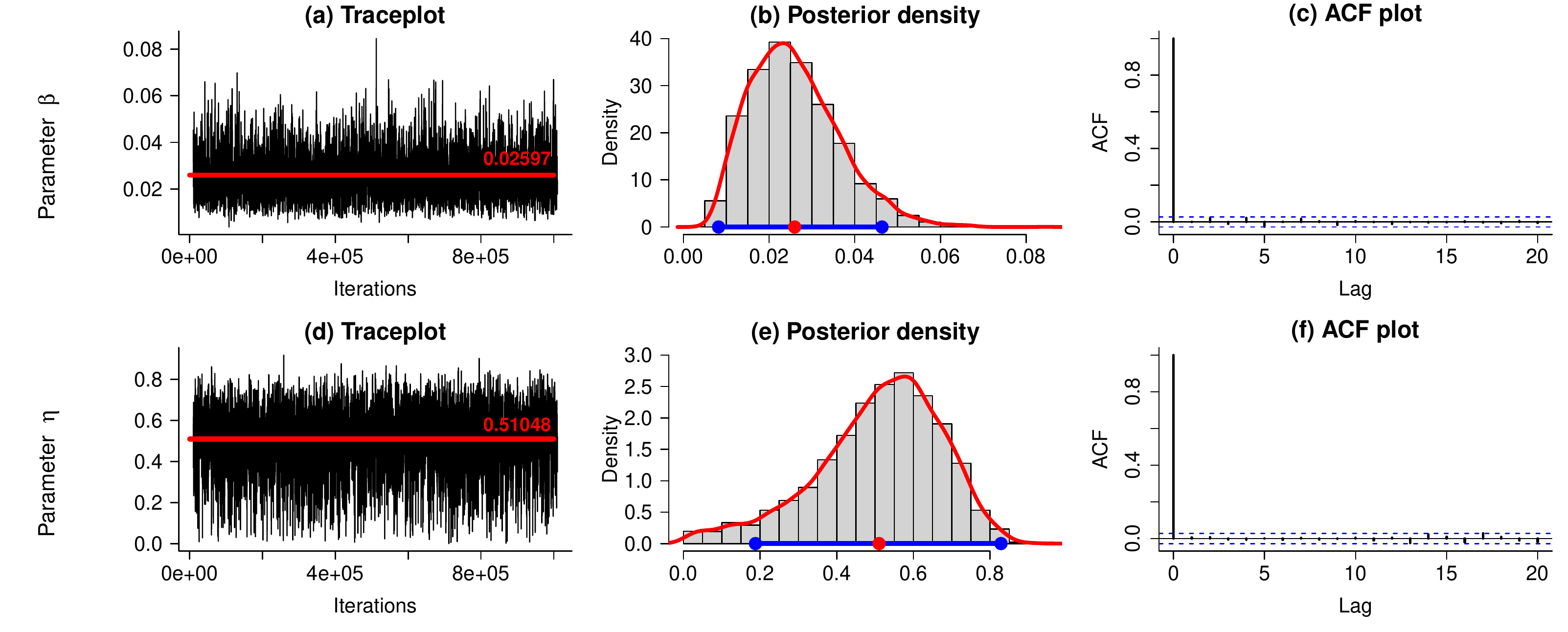}
	\caption[figDES]{Posterior samples for the model parameters of the DB distribution applied to data from patients with pelvic tumors. (a) Traceplots of posterior samples, (b) histograms and posterior densities with the correspondent $95\%$ HDI (blue lines), and (c) auto-correlation function (ACF) plots for the posterior samples of the model parameters.\bigskip}
	\label{fig:FigBayesEx03}
\end{figure}
Figure \ref{fig:FigResidualsEx03} describes the randomized quantile residuals from the fitted models. The model based on the DB distribution appears to fit the data reasonably well, as does the models based on the DsFx-I and the Lindley distribution. Models based on the discrete Rayleigh and discrete Burr-Hatke distributions did not fit the data well.

Assuming the Bayesian model introduced in Section \ref{Bayes}, we assumed prior distributions $\beta \sim Gamma(0.001,0.001)$ and $\eta \sim Beta(1,1)$, that are approximately non-informative priors for the model parameters. Posterior means for $\beta$ and $\eta$ are $\widehat{\beta }_{Bayes} = 0.02597$ (95\% HDI 0.0082 to 0.0464) and $\widehat{\eta }_{Bayes} = 0.51048$ (95\% HDI 0.1879 to 0.8286), respectively. Figure \ref{fig:FigBayesEx03} describes the posterior samples for the parameters $\beta$ and $\eta$. Panels (a) and (d) show that the MCMC chains reached satisfactory convergence for both parameters and the Geweke Z scores for $\beta$ and $\eta$ chains are -0.060 and -0.729, respectively. Panels (b) and (e) describe the posterior densities and the 95\% HDI. Panels (c) and (f) shows that autocorrelations within each chain were reasonably low.

\section{Concluding Remarks} \label{CR}
The literature contains few articles on the DB distribution introduced by (\citeauthor{Altun2020}, \citeyear{Altun2020}). Hence the contribution of the present article is the introduction of parameter estimation for DB distribution considering the inclusion of right-censored data and a cure fraction. Three applications to real datasets show that the model based on DB distribution performs at least as good as some other traditional discrete models as the DsFx-I, discrete Lindley, discrete Rayleigh, and discrete Burr-Hatke distributions. Therefore, the model based on DB distribution showed to be a suitable way to analyze discrete survival data, even including the presence of immune individuals. Moreover, the model can be easily implemented in computational programs as R, as showed in the Appendix. Currently, we find many examples of application of cure rate models to medical data (\citeauthor{Gallardo2021}, \citeyear{Gallardo2021}; \citeauthor{Leao2020}, \citeyear{Leao2020}; \citeauthor{Rafati2020}, \citeyear{Rafati2020}), which makes these models attractive to be assumed in lifetime data analysis. The methods introduced in this paper could be very helpful to researchers dealing with discrete survival data.

\section*{Appendice: R Codes}

Under the frequentist approach, the following R code is used to implement the model for survival data with a cure fraction based on the DB distribution, as presented in subsection \ref{cure}. We used the function maxLik of the maxLik package (\citeauthor{Henningsen2011}, \citeyear{Henningsen2011}) for the maximization of the likelihood function.
\begin{verbatim}
	# Reading data (Wang et al., 2015)
	t <- c(3,7,11,18,22,25,28,32,34,35,35,36,40,40,41,54,66,76,84,88,92)
	d <- c(1,1,0,1,0,1,1,0,0,1,0,0,0,0,1,0,0,0,0,0,0)
	n <- length(t)  # the sample size
	K <- 2          # number of parameters
	
	# Loading the maxLik package
	library(maxLik)
	# The likelihood function
	log.f <- function(parms) {
		beta  <- parms[1]
		eta   <- parms[2]
		if (parms[1]<0) return(-Inf)
		if (parms[2]<0) return(-Inf)
		if (parms[2]>1) return(-Inf)
		p    <- exp(-beta)
		St0  <- (3-2*p^(t+1))*p^(2*(t+1))
		ft0  <- p^(2*t)*(p-1)*(2*p^t*(p^2+p+1)-3*p-3)
		St   <- eta + (1-eta)*St0
		ft   <- (1-eta)*ft0
		like <- ft^d * St^(1-d)
		L    <- sum(log(like))
		if (is.na(L)==TRUE) {return(-Inf)} else {return(L)} }
	# Obtaining the ML estimates
	mle  <- c()
	mle  <- maxLik(logLik=log.f,start=c(0.08,0.6))
	summary(mle)
	betaDB <-mle$estimate[1]
	etaDB  <-mle$estimate[2]
	s <- vcov(mle)
	# The 95% confidence intervals
	llimDB  <- round(betaDB - qnorm(0.975) * sqrt(s[1,1]),4)
	ulimDB  <- round(betaDB + qnorm(0.975) * sqrt(s[1,1]),4)
	llimDBe <- round(etaDB  - qnorm(0.975) * sqrt(s[2,2]),4)
	ulimDBe <- round(etaDB  + qnorm(0.975) * sqrt(s[2,2]),4)
	cat("n = ",n,"\n")
	cat("Beta  = ",betaDB, "95%CI: (",llimDB,",",ulimDB, ") \n")
	cat("Eta   = ",etaDB,  "95%CI: (",llimDBe,",",ulimDBe, ") \n")
	# Calculating AIC, BIC and AICC
	aic  <- AIC(mle)
	bic  <- AIC(mle,k = log(n))
	aicc <- aic + (2*K^2+2*K)/(n-K-1)
	cat("AIC = ",aic,", BIC = ",bic,", AICC = ",aicc,"\n")
\end{verbatim}

This is the R code for the Bayesian model for survival data with a cure fraction based on the DB distribution, as presented in subsection \ref{Bayes}:
\begin{verbatim}
	# The log posterior function
	log.post <- function(t,d,parms) {
		beta    <- parms[1]
		eta     <- parms[2]
		if (parms[1]<0) return(-Inf)
		if (parms[2]<0) return(-Inf)
		if (parms[2]>1) return(-Inf)
		p    <- exp(-beta)
		St0  <- (3-2*p^(t+1))*p^(2*(t+1))
		ft0  <- p^(2*t)*(p-1)*(2*p^t*(p^2+p+1)-3*p-3)
		St   <- eta + (1-eta)*St0
		ft   <- (1-eta)*ft0
		like <- ft^d * St^(1-d)
		log.like <- sum(log(like))
		prior    <- dgamma(beta,0.001,0.001)*dbeta(eta,1,1)
		log.prior <- log(prior)
		L <- log.like + log.prior
		if (is.na(L)==TRUE) {return(-Inf)} else {return(L)} }
	
	# Obtaining the MCMC estimates
	posterior <- MCMCmetrop1R(log.post,theta.init=c(beta=0.05,eta=0.6), 
	burnin=10000, mcmc=1000000, thin=200, logfun=T, t=t, d=d, verbose=100000,
	tune = 1)
	varnames(posterior) <- c("beta","eta")
	summary(posterior)
	# Obtaining the HPD intervals
	HPDinterval(posterior, prob = 0.95)
	# Geweke z scores
	geweke.diag(posterior)
\end{verbatim}

\end{document}